\documentclass[prl,twocolumn,lengthcheck,superscriptaddress]{revtex4-1}

\usepackage{graphicx,color}
\usepackage{amsmath,amssymb}
\usepackage{textcomp}
\usepackage{braket,slashed}
\usepackage{hyperref}
\usepackage[english]{babel}

\usepackage{bm}

\usepackage[mathscr]{euscript}
\usepackage{dsfont}

\newcommand{\operator}[1]{\ensuremath{\hat{#1}}}
\newcommand{\voperator}[1]{\ensuremath{\mathds{#1}}}
\newcommand{\manifold}[1]{\ensuremath{\mathcal{#1}}}

\newcommand{\vectorspace}[1]{\ensuremath{\mathbb{#1}}}

\DeclareMathOperator{\order}{\mathscr{O}}
\DeclareMathOperator{\tr}{tr}

\newcommand{\ic}{\ensuremath{\mathrm{i}}}

\newcommand{\rket}[1]{\ensuremath{|#1)}}
\newcommand{\rbra}[1]{\ensuremath{(#1|}}
\newcommand{\rbraket}[1]{\ensuremath{(#1)}}

\newcommand{\hilbert}{\ensuremath{\vectorspace{H}}}

\newcommand{\varM}{\ensuremath{\manifold{M}}}

\newtheorem{alg}{Algorithm}

\begin{document}
\title{Unifying time evolution and optimization with matrix product states}

\author{Jutho Haegeman}
\affiliation{Department of Physics and Astronomy, University of Ghent, Krijgslaan 281 S9, B-9000 Ghent, Belgium}
\author{Christian Lubich}
\affiliation{Mathematisches Institut, Universit\"{a}t T\"{u}bingen, Auf der Morgenstelle 10, D-72076 T\"{u}bingen, Germany}
\author{Ivan Oseledets}
\affiliation{Skolkovo Institute of Science and Technology, Novaya St.~100, 143025 Skolkovo, Russia}
\affiliation{Institute of Numerical Mathematics, Russian Academy of Sciences, Gubkina Street, 8, Moscow, Russia}
\author{Bart Vandereycken}
\affiliation{Department of Mathematics, Princeton University, Fine Hall, Princeton, NJ 08544, USA}
\author{Frank Verstraete}
\affiliation{Department of Physics and Astronomy, University of Ghent, Krijgslaan 281 S9, B-9000 Ghent, Belgium}
\affiliation{Faculty of Physics, University of Vienna, Boltzmanngasse 5, A-1090 Wien, Austria}

\begin{abstract}
We show that the time-dependent variational principle provides a unifying framework for time-evolution methods and optimisation methods in the context of matrix product states. In particular, we introduce a new integration scheme for studying time-evolution, which can cope with arbitrary Hamiltonians, including those with long-range interactions. Rather than a Suzuki-Trotter splitting of the Hamiltonian, which is the idea behind the adaptive time-dependent density matrix renormalization group method or time-evolving block decimation, our method is based on splitting the projector onto the matrix product state tangent space as it appears in the Dirac-Frenkel time-dependent variational principle. We discuss how the resulting algorithm resembles the density matrix renormalization group (DMRG) algorithm for finding ground states so closely that it can be implemented by changing just a few lines of code and it inherits the same stability and efficiency. In particular, our method is compatible with any Hamiltonian for which DMRG can be implemented efficiently and DMRG is obtained as a special case of imaginary time evolution with infinite time step.
\end{abstract}

\maketitle
Tensor network states, and in particular matrix product states (MPS) \cite{1992CMaPh.144..443F,1995PhRvL..75.3537O,2008AdPhy..57..143V,2009JPhA...42X4004C}, have become increasingly popular and successful for the description of strongly interacting quantum many body systems. While originating from the efficient and robust density matrix renormalization group (DMRG) algorithm \cite{1992PhRvL..69.2863W,1993PhRvB..4810345W} for finding ground states of one-dimensional (and quasi two-dimensional) spin systems, an undeniable influence for this success has also been the ability to accurately study dynamical properties using time-evolution methods such as the time-evolving block decimation (TEBD) method \cite{2004PhRvL..93d0502V,2004PhRvL..93g6401W,2004JSMTE..04..005D,2004PhRvL..93t7204V}. The reformulation of DMRG in terms of MPS \cite{1995PhRvL..75.3537O,1997PhRvB..55.2164R,2004PhRvL..93v7205V,2011AnPhy.326...96S} has been of key importance in this development. Being able to probe and understand quantum dynamics is becoming increasingly valuable for experiments at low temperature or high energy, or for theoretical questions such as thermalization. Nevertheless, quantum dynamics is mostly inaccessible to alternative methods such as Monte Carlo sampling of the partition function. 

The key ingredient of TEBD-based methods is a Lie-Trotter-Suzuki \cite{Trotter:1959aa,Suzuki:1976aa} splitting of the Schr\"{o}dinger equation according to the individual terms $\operator{h}_i$ of the Hamiltonian $\operator{H}=\sum_{i} \operator{h}_i$. When these terms are local, they can be applied efficiently and the tensor network can be updated accordingly. The resulting increase of the virtual dimensions of the tensors can be countered by a subsequent truncation step, although the growth of entanglement entropy under time evolution indicates that a net increase of the dimensions is inevitable if an accurate description is required. Unfortunately, the TEBD idea is not applicable to long-range interactions, which appear when using MPS for quasi-2D systems or quantum chemistry applications.

An alternative idea for time evolution, the Dirac-Frenkel time-dependent variational principle (TDVP), was formulated only recently for the variational class of MPS \cite{2011PhRvL.107g0601H,LubRSV:2013}. The key ingredient is to project the right hand side of the Schr\"{o}dinger equation, $\operator{H}\ket{\Psi}$, onto the tangent space, so that the evolution never leaves the manifold. This approach is independent of the Hamiltonian and can be implemented efficiently for long-range Hamiltonians. The TDVP equations define a simultaneous update for all MPS tensors and, as a consequence thereof, its original implementation depends on inverses of matrices conditioned by the Schmidt coefficients, similar to the original TEBD implementation. This results in the paradoxical situation that the stability of this method deteriorates as the approximation of the exact state (corresponding to the value of the smallest Schmidt coefficient) is improved. More concretely, the TDVP equations suffer from what is known as stiffness in the numerical analysis literature \cite{wanner1991solving} and this results in a severe stepsize restriction when using standard explicit time integrators. Furthermore, the TDVP algorithm has so far not gained widespread acceptance and was in fact criticised to be complex and distinct from DMRG-inspired algorithms in Ref.~\onlinecite{2014arXiv1407.1832Z}, where another algorithm for time-evolution with long-range interactions is proposed, based on an approximation of the evolution operator in terms of a matrix product operator (MPO) \cite{2007JSMTE..10...14M,2010NJPh...12b5012P}.

This Letter overcomes such criticism by presenting an alternative integration scheme for the TDVP equations for finite MPS, based on a Lie-Trotter splitting of the tangent space projector. The resulting algorithm resembles DMRG so closely that it can be obtained by merely modifying 2 lines of code, and the one-site DMRG method is obtained as special case of imaginary time evolution. In particular, this time evolution algorithm is compatible with any Hamiltonian for which ground state DMRG can be executed efficiently, e.g. long-range Hamiltonians which are written in the form of MPOs. We then discuss how to implement a 2-site scheme that allows to dynamically increase the bond dimension. We also relate our approach to the more conventional time-dependent DMRG or TEBD methods, which are restricted to local (typically nearest neighbor) interactions. The resulting ideas will also be relevant for continuous MPS (cMPS) \cite{2010PhRvL.104s0405V,2013PhRvB..88h5118H}, which seem incompatible with traditional DMRG- and TEBD-based approaches.

\begin{figure*}
\includegraphics[width=\textwidth]{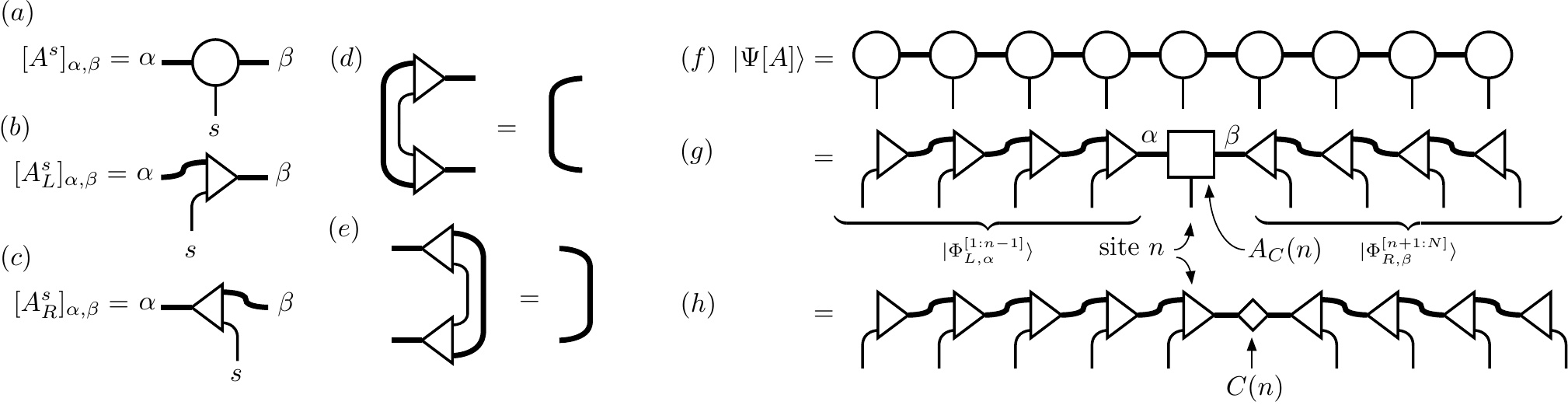}
\caption{An arbitrary MPS tensor is represented as a shape with fat legs corresponding to the virtual indices of dimension $D$ and a normal leg corresponding to the physical index of dimension $d$ (a). The left-orthonormal [right-orthonormal] MPS tensors are represented using triangles (b) [(c)] and satisfies the condition (d) [(e)]. An MPS $\ket{\Psi[A]}$ using general tensors (f) can be brought into a mixed-canonical form with a one-site center $A_C(n)$ at site $n$ (g) or even a zero-site center $C(n)$ (h), whose singular values correspond to the Schmidt coefficients of the state.}
\label{fig:defs}
\end{figure*}

We define a matrix product state (MPS) with open boundary conditions, also known as a tensor train \cite{oseledets2011tensor} in the numerical mathematics community, as
\begin{displaymath}
\ket{\psi[A]}=\sum_{\{s_{n}\}=1}^{d} A^{s_{1}}(1) A^{s_{2}}(2) \cdots A^{s_{N}}(N) \ket{s_{1}s_{2}\ldots s_{N}}
\end{displaymath}
where we used square brackets $[]$ to denote the dependence on \emph{a set of} site-dependent matrices $A^{s}(n)$ having site-dependent dimensions $D_{n-1}\times D_{n}$. As the whole matrix product has to evaluate to a scalar, we have $D_0 = D_N=1$. A possible site-dependence for the dimension $d$ of the local Hilbert space is not explicitly denoted for the sake of brevity. To introduce the required concepts, we first summarize the finite-size DMRG method algorithm in the MPS framework and refer to Fig.~\ref{fig:defs} and the excellent review of Ref.~\onlinecite{2011AnPhy.326...96S} for further details.

A key ingredient is the observation that the physical state $\ket{\Psi[A]}$ is unchanged under the gauge transformation $A^s(n)\mapsto A_G^s(n)=G(n-1)^{-1} A^s(n) G(n)$. This gauge freedom can be used to impose certain canonical formats on the MPS representation, in terms of left-orthonormal matrices $A_L^s(n)$ or right-orthonormal matrices $A_R^s(n)$ [see Fig.~\ref{fig:defs}(b-e)]. The stability of finite MPS methods originates from the fact that we can transform the original tensors $A^s(n)$ into e.g.\ the left orthornormal representation without explicit computation of the inverses $G(n)^{-1}$. Starting from $C(0)=1$, an orthogonal factorization [e.g. QR-decomposition] allows to write $C(n-1) A^s(n) = A_L^s(n) C(n)$. The final $C(N)$ is a scalar representing the norm of the state, and can be discarded. If the original MPS was already in the right canonical form, the singular values of $C(n)$ correspond to the Schmidt coefficients of $\ket{\Psi}$ for a bipartition of the lattice in $[1:n]$ and $[n+1:N]$. Defining the one-site center block $A^s_C(n)=C(n-1) A_R^s(n) = A_L^s(n) C(n)$ allows to write the state in the mixed canonical form [see also Fig.~\ref{fig:defs}(g)]
\begin{equation}
\ket{\Psi}=\sum_{\alpha,s_{n},\beta} [A^{s_n}_C(n)]_{\alpha,\beta} \ket{\Phi^{[1:n-1]}_{L,\alpha}} \ket{s_n} \ket{\Phi^{[n+1:N]}_{R,\beta}}\label{eq:defphi}
\end{equation}
where new states $\ket{\Phi^{[1:n-1]}_{L,\alpha}}$ ($\ket{\Phi^{[n+1:N]}_{R,\beta}}$) constitute an orthonormal basis for the left (right) block of the lattice.

\begin{figure*}
\includegraphics[width=\textwidth]{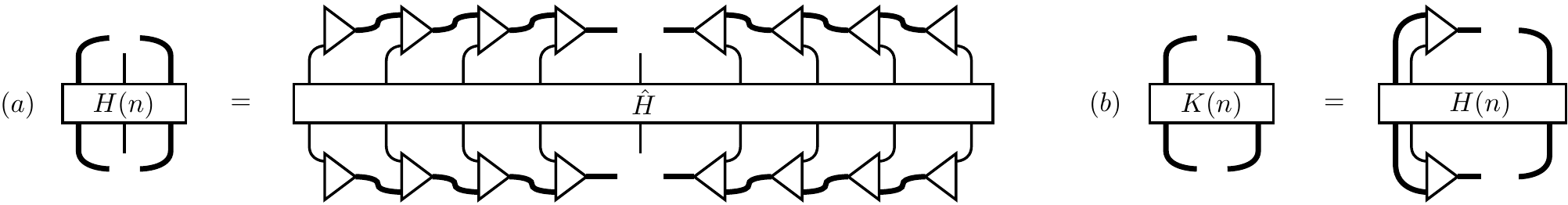}
\caption{(a) One-site effective Hamiltonian $H(n)$. (b) Similarly, one can define a zero-site effective Hamiltonian $K(n)$ which can easily be computed from $H(n)$.}
\label{fig:effham}
\includegraphics[width=\textwidth]{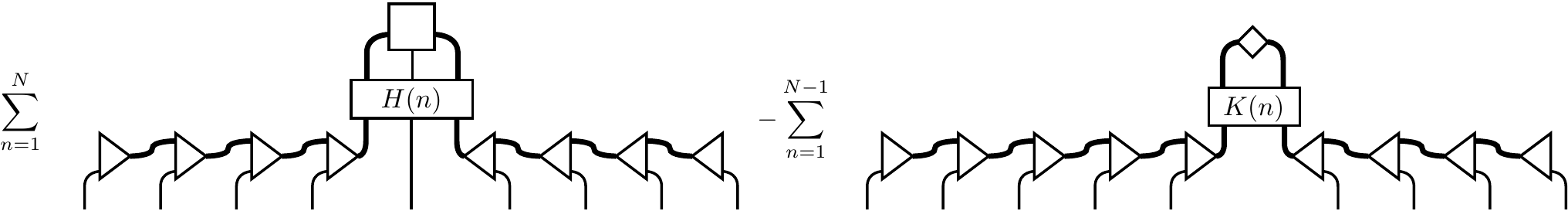}
\caption{Right-hand side (up to the factor $-i$) of the TDVP equation [Eq.~\eqref{eq:TDVPgeom}].}
\label{fig:tdvp}
\end{figure*}
The one-site DMRG method in MPS language \cite{2011AnPhy.326...96S}, where a single tensor is optimized, corresponds to an alternating least squares optimization for a homogeneous problem and typically results in a generalized eigenvalue problem. By gauging the MPS such that the current tensor being optimized is the center site $A^s_C(n)$, this is transformed into a standard eigenvalue problem with an effective one-site Hamiltonian $H(n)$ defined in Fig.~\ref{fig:effham}(a). The matrix vector product of $H(n)$ can be computed efficiently for typical physical Hamiltonians having either short-ranged interactions or an MPO form, so that an iterative eigensolver can be used. After updating $A_C^s(n)$ as the lowest eigenvector of $H(n)$, one performs the orthogonal factorization to shift the center site to the left or right, so that the next tensor can be optimized. Reaching one end of the chain, one can then sweep back, so that every orthogonalization step is immediately followed by an optimization step. For further reference, we also define a zero-site effective Hamiltonian $K(n)$ [see Fig.~\ref{fig:effham}(b)] for the degrees of freedom $C(n)$ on the virtual bond between sites $n$ and $n+1$, as in [see also Fig.~\ref{fig:defs}(h)]
\begin{equation}
\ket{\Psi}=\sum_{\alpha,\beta} [C(n)]_{\alpha,\beta} \ket{\Phi^{[1:n]}_{L,\alpha}} \ket{\Phi^{[n+1:N]}_{R,\beta}}.\label{eq:statewithC}
\end{equation}

Geometrically, the TDVP corresponds to an orthogonal projection of the evolution vector of the Schr\"{o}dinger equation ($-\ic H \ket{\Psi}$) onto the tangent space of the MPS manifold $\varM_{\text{MPS}}$ at the current state
\begin{equation}
\frac{\mathrm{d} \ket{\Psi(A)}}{\mathrm{d} t} = -\ic \operator{P}_{T_{\ket{\Psi(A)}} \varM_{\text{MPS}}} \operator{H}\ket{\Psi(A)}\label{eq:TDVPgeom}
\end{equation}
which has the required effect that the time-evolving state will never leave the MPS manifold and can thus be described by time-evolving parameters as $\ket{\Psi(A(t))}$. The TDVP [Eq.~\eqref{eq:TDVPgeom}] is a complicated set of non-linear equations for all degrees of freedom in the MPS. The algorithm presented in Ref.~\onlinecite{2011PhRvL.107g0601H} used the explicit Euler integration scheme to simultaneously update all parameters, which was only possible by using explicit inverses of matrices containing very small singular values. However, it is a straightforward calculation [see Ref.~\onlinecite{2014arXiv1407.2042L} and supplementary material] to show that the tangent space projector $\operator{P}_{T_{\ket{\Psi(A)}} \varM_{\text{MPS}}}$ satisfies
\begin{equation}
\sum_{n=1}^{N} \operator{P}_{L}^{[1:n-1]}\otimes \operator{1}_n\otimes \operator{P}_{R}^{[n+1:N]}-\sum_{n=1}^{N-1} \operator{P}_{L}^{[1:n]}\otimes \operator{P}_{R}^{[n+1:N]}
\label{eq:projsplitting}
\end{equation}
where $\operator{P}_{L}^{[1:n]}=\sum_{\alpha=1}^{D} \ket{\Phi^{[1:n]}_{L,\alpha}}\bra{\Phi^{[1:n]}_{L,\alpha}}$ and $\operator{P}_{R}^{[n:N]}=\sum_{\beta=1}^{D} \ket{\Phi^{[n:N]}_{R,\beta}}\bra{\Phi^{[n:N]}_{R,\beta}}$. These operators are independent of the unitary freedom that remains in the definition of $\ket{\Phi^{[1:n]}_{L,\alpha}}$ and $\ket{\Phi^{[n:N]}_{R,\beta}}$ and correspond to orthogonal projector onto the support of the reduced density matrix $\rho$ of the state $\ket{\Psi[A]}$ in the corresponding regions; see also Fig.~\ref{fig:tdvp}.

The new insight is that every single term out of Eq.~\eqref{eq:projsplitting} can be integrated exactly \cite{2014arXiv1407.2042L}. For example, using the single projector $\operator{P}_{L}^{[1:n-1]}\otimes \operator{1}_n\otimes \operator{P}_{R}^{[n+1:N]}$ and writing $\ket{\Psi[A]}$ as in Eq.~\eqref{eq:defphi} with center site $n$, the exact solution can be obtained by only making $A_C(n)$ time-dependent and letting it satisfy $\dot{\bm{A}}_C(n,t)=-\ic H(n) \bm{A}_C(n,t)$ where a bold notation is used for the vector representation of $A_C(n)$. We can thus explicitly integrate this differential equation by only making $A_C(n)$ time dependent as
\begin{equation}
\bm{A}_C(n,t)=\exp\big[-\ic H(n) t\big] \bm{A}_{C}(n,0).\label{eq:evolveAC}
\end{equation}
Similarly, a projector term of the form $-\operator{P}_{L}^{[1:n]}\otimes \operator{P}_{R}^{[n+1:N]}$ can be integrated explicitly by writing the $\ket{\Psi}$ as in Eq.~\eqref{eq:statewithC} and making $C(n)$ time-dependent as
\begin{equation}
\bm{C}(n,t)=\exp\big[+\ic K(n) t \big] \bm{C}(n,0).\label{eq:evolveC}
\end{equation}
Note that opposite sign in the exponentials of Eq.~\eqref{eq:evolveAC} and Eq.~\eqref{eq:evolveC}. The evolution of $C$ in Eq.~\eqref{eq:evolveC} can be interpreted as an evolution backwards in time. Similar to the eigenvalue problem in DMRG, both Eq.~\eqref{eq:evolveAC} and \eqref{eq:evolveC} can be integrated using an iterative Lanczos scheme \cite{doi:10.1137/S0036142995280572} in order to keep the computational cost at $\mathcal{O}(D^3)$, but this is a well-controlled approximation for which the error can be made arbitrarily small. We refer to Ref.~\onlinecite{Lubich:2008fk} for estimators for the error resulting from the Lanczos exponentiation.

For this type of differential equation consisting of a sum of integrable parts, it is very natural to use a Lie-Trotter splitting approach \cite{Trotter:1959aa,Hairer:2004aa,Lubich:2008fk}, where one evolves according to each integrable part for a small time $\Delta t$. There is a natural order similar to the DRMG sweep, which optimizes the computational overhead (and results in robustness in the case of overapproximation \cite{2014arXiv1407.2042L}). For a right orthogonal MPS, start at $n=1$ and repeat the following steps: Evolve $A_C(n)$ according Eq.~\eqref{eq:evolveAC} for a time step $\Delta t$. Factorize the updated $A^s_C(n)=A_L^s(n) C(n)$. Evolve $C(n)$ backwards in time according to Eq.~\eqref{eq:evolveC} before absorbing it into the next site to create $A^s_C(n+1)=C(n) A_R^s(n+1)$. Having a Lanczos routine for computing the matrix exponential acting on a vector (e.g.~Ref.~\onlinecite{sidje1998expokit}), this algorithm differs from one-site DMRG by replacing the optimization step for $A_C(n)$ (eigenvalue solver) with the evolution step, and by adding an extra evolution step for $C(n)$ before absorbing it at the next site. Completing a single left-to-right sweep is a first-order integrator that produces an updated $\ket{\Psi}$ at time $t+\Delta t$ with a local integration error of order $\mathcal{O}(\Delta t^2)$. Completing the right-to-left sweep is equivalent to composing this integrator with its adjoint, resulting in a second-order symmetric method \footnote{A symmetric integration scheme has the property that, applying it with time step $\Delta t$ and then with time step $-\Delta t$, the initial state is recovered} so that the state at time $t+2\Delta t$ has a more favourable error of order $\mathcal{O}(\Delta t^3)$. It is thus natural to set $\Delta t\to\Delta t/2$ and to define the complete sweep (left and right) as a single integration step. It is also possible to obtain higher order integrators by applying composition schemes to this symmetric integrator \cite{Hairer:2004aa}. Note that the finite time step errors refer to errors with respect to the exact solution of the TDVP differential equation. The error with respect to the original Schr\"{o}dinger equation also receives a contribution from the discrepancy between the TDVP evolution and the Schr\"{o}dinger evolution. As explained in Ref.~\onlinecite{Lubich:2008fk}, this source of error can be bound in terms of $\lVert (\operator{1}-\operator{P}_{T_{\ket{\Psi[A]}}\varM_{\text{MPS}}})\operator{H}\ket{\Psi[A]}\rVert$, a quantity that can easily be evaluated \cite{PhysRevB.88.075133}.

This integration scheme contains evolution for the 1-site and 0-site center blocks $A_C(n)$ and $C(n)$ according to the respective effective Hamiltonian matrices $H(n)$ or $K(n)$, which are Hermitian. Therefore, the resulting evolution will have exact norm and energy conservation (in case of a time-independent Hamiltonian). This integration scheme can also be combined with imaginary time evolution $t\mapsto -\ic \tau$, e.g.\ for constructing (purifications of) thermal states or for finding ground states. In the latter case, the goal is not to accurately capture the full evolution and the scheme can be varied \footnote{For example, the backward evolution of $C(n)$, which actually increases the energy but potentially allows to avoid local minima, can be omitted.}. In the particular case of $\Delta \tau\to \infty$, the evolution of $A_C(n)$ will actually force it to be the lowest eigenvector of $H(n)$, exactly as in the optimization step of the one-site DMRG algorithm. One can check that the resulting $C(n)$ is then an exact eigenstate of $K(n)$, so that its backwards evolution has no further effect. Hence, this method becomes identical to 1-site DMRG.

To implement 2-site algorithms in this framework, the tangent space projector in Eq.~\eqref{eq:TDVPgeom} has to be replaced by a projector onto the linear space of 2-site variations, which is spanned by the states $\sum_{n=1}^{N-1} \sum_{\{s_{n}\}=1}^{d} A^{s_{1}}(1) \cdots A^{s_{n-1}}(n-1) B^{s_{n}s_{n+1}}(n:n+1) A^{s_{n+2}}(n+2)\cdots A^{s_{N}}(N)\ket{s_{1}\ldots s_{n}\ldots s_{N}}$. It is now tempting to propose the following 2-site integration scheme: evolve a 2-site center block $A_C(n:n+1)$ according to its effective Hamiltonian $H(n:n+1)$, factor it into $A_C^{s_n,s_{n+1}}(n:n+1)\to A_L^{s_n}(n) A_C^{s_{n+1}}(n+1)$ and evolve $A_C^{s_{n+1}}(n+1)$ backwards in time according to $H(n+1)$ before absorbing it in the next 2-site block $A_C^{s_{n+1}}(n+1) A_C^{s_{n+2}}(n+2)\to A_C^{s_{n+1},s_{n+2}}(n+1:n+2)$. We validate this approach in the supplementary material. Every 2-site block that is factored can increase the local bond dimension. This requires a local truncation based on the singular values in every step or a global approximation with the best MPS with smaller bond dimensions at the end of the sweep \cite{2004PhRvL..93t7204V}. It is straightforward to show that imaginary time evolution with a time step $\Delta \tau\to\infty$ gives rise to 2-site DMRG. The TEBD algorithm also fits in this framework, by observing that for nearest-neighbor interactions, the evolution vector $\hat{H}\ket{\Psi}$ fits exactly in the space of 2-site variations (no projection is necessary) by choosing e.g. $B^{s_n,s_{n+1}}(n:n+1)=\sum_{t_n,t_{n+1}} \braket{s_n,s_{n+1}|\hat{h}_{n,n+1}|t_{n},t_{n+1}}A^{t_{n}}(n) A^{t_{n+1}}(n+1)$ and using a Trotter splitting based on these terms. The representation $B^{s_n,s_{n+1}}(n:n+1)$ is however non-unique, and our algorithm based on splitting the projector (even when it acts trivially) gives rise to a slightly different scheme. The error analysis of the 2-site algorithms is more complicated and is discussed in more detail in the supplementary material.

\begin{figure}
\includegraphics[width=\columnwidth]{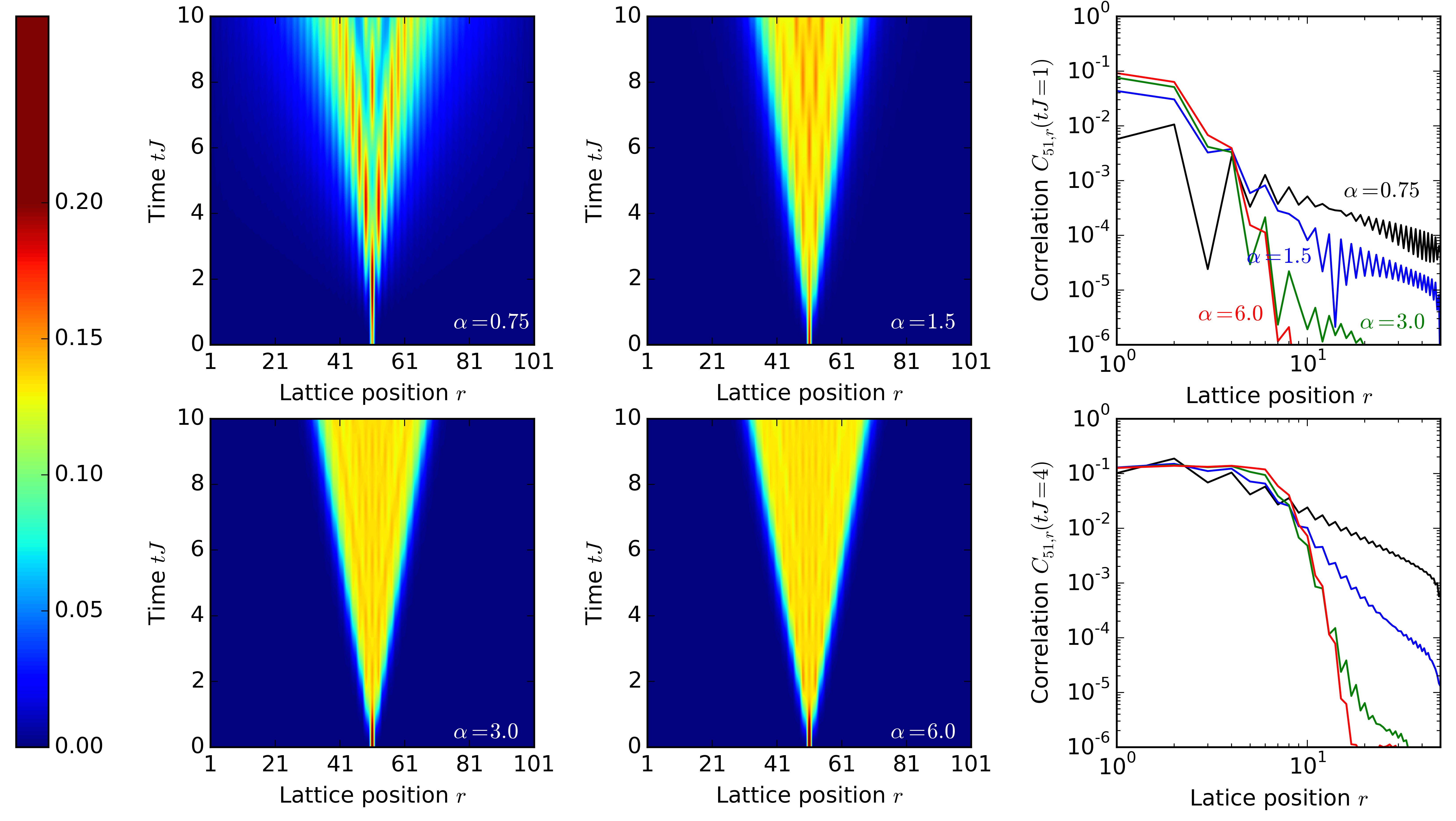}
\caption{Absolute value of $\braket{\Psi_{0}|U^\dagger \sigma^x_r(t) U|\Psi_{0}}-\braket{\Psi_{0}|\sigma^x_r(t) |\Psi_{0}}$ with $H$ the XY Hamiltonian in Eq.~\eqref{eq:hamxy}, $\ket{\Psi_{0}}$ the ground state of $H$ and $U=\exp(\ic \sigma^y_{51} \pi/4)$ on a chain of $N=101$ spins, as function of $r$ and $t$ for various values of $\alpha$.}
\label{fig:xy}
\end{figure}

As an application, we study a one-dimensional model with power-law decay of interactions. There is recently a strong interest in such models as they can be realized in experiments with ultra-cold matter, and it is thus actively investigated in what regimes they still exhibit a light cone and what the corresponding shape of that light cone is \cite{PhysRevLett.111.207202,2014PhRvL.113c0602G}. We consider the particular case of the XY model, given by the Hamiltonian
\begin{equation}
H=\frac{1}{2}\sum_{i<j=1}^{N} \frac{J}{\lvert i-j\rvert^{\alpha}} \left(\sigma^x_i\sigma^x_j + \sigma^y_i \sigma^y_j\right)\label{eq:hamxy}
\end{equation}
on a finite chain of $N$ sites with open boundary conditions. This Hamiltonian was recently realized in an experiment with trapped ${}^{171}\mathrm{Yb}^{+}$ ions \cite{Richerme:2014aa}. 

In Ref.~\onlinecite{2014PhRvL.113c0602G}, the time dependent expectation value $\sigma_r^x(t)$ was compared between a fully polarized product state $\ket{\Psi_0}=\otimes_{n} \ket{\uparrow}_n$ and the perturbed state $U\ket{\Psi_0}$ with $U=\exp(\ic \pi \sigma^y_0 /4)$ as function of $r$ and $t$ for various values $\alpha>1$. For this choice of $\ket{\Psi_0}$, the quantum many body problem can be reduced to a single particule problem. In accordance with the general result of that paper, there is a light cone and the leakage outside the light cone contains a contribution with exponential decay that is determined by the nearest neighbor interactions and is thus independent of $\alpha$, as well as a contribution with power law decay that dominates at larger distances. We repeat this set-up but using the fully correlated ground state of $H$ as $\ket{\Psi_0}$, approximated as MPS. We simulate the time evolution with our 2-site integrator using a 4th order composition method using a time step $\mathrm{d}t=0.02/J$ for an open chain with $N=101$ sites (Fig.~\ref{fig:xy}). The long-range interactions of the Hamiltonian were approximated by an MPO with a maximal absolute error smaller than $10^{-8}$. We also consider a value of $\alpha=0.75<1$, for which the results from Ref.~\onlinecite{2014PhRvL.113c0602G} are not applicable. For $\alpha>1$, we observe a light cone and the leakage outside is in perfect agreement with the theoretical predictions of Ref.~\onlinecite{2014PhRvL.113c0602G}. The light cone itself seems to be linear for $\alpha=3.0$ and $\alpha=6.0$, but the shape is less clear for $\alpha=1.5$. For $\alpha=0.75$ there is no sharp wave front but it still seems to take a finite amount of time to signal over a finite distance, up to power law corrections.

In conclusion, we have formulated a time integration scheme for the TDVP in the context of MPS, which is based on a Trotter decomposition of the tangent space projector rather than the Hamiltonian terms. This allows to simulate time evolution using an algorithm that is remarkably similar to DMRG and can deal with arbitrary long-range Hamiltonians. This approach will also be relevant for formulating new algorithms to deal with continuous MPS \cite{2010PhRvL.104s0405V,2013PhRvB..88h5118H}, where algorithms based on a Trotter decomposition of the Hamiltonian terms are not applicable.

\acknowledgements{JH acknowledges the invitation by CL to the Universit\"{a}t T\"{u}bingen, where this research was initiated. We thank Z.~Gong, M.~Foss-Feig and A.~Gorshkov for the suggestion to study the long-range XY model. This work was supported by Russian Science Foundation grant 14-11-00659 (IO), by DFG projects SPP 1324 and GRK 1838 (CL), by the Austrian FWF SFB grants FoQuS and ViCoM and the European grants SISQ and QUTE.}

\appendix
\begin{widetext}
\section*{Supplementary material for\\
``Unifying time evolution and optimization with matrix product states''}

\subsection{Derivation of the tangent space projector}
To derive the main result of this paper, we have to introduce some further notation. We repeat the definition of a general matrix product state with open boundary conditions, given by
\begin{equation}
\ket{\psi[A]}=\sum_{\{s_{n}\}=1}^{d} A^{s_{1}}(1) A^{s_{2}}(2) \cdots A^{s_{N}}(N) \ket{s_{1}s_{2}\ldots s_{N}}.\label{eq:mps}
\end{equation}
If we want to compute expectation values without bringing the MPS in a specific format, we can first define the sets of site-dependent $D_{n}\times D_{n}$ density matrices $l(n)$ and $r(n)$ (with $n=0,\ldots,N$) for the auxiliary system through $l(0)=1=r(N)$ and
\begin{align}
l(n)&=\sum_{s=1}^{d} (A^{s}(n))^{\dagger} l(n-1) A^{s}(n),\\
r(n)&=\sum_{s=1}^{d} A^{s}(n+1) r(n+1) (A^{s}(n+1))^{\dagger}.
\end{align}
The norm of the state is then given by $r(0)=l(N)=\rbraket{l(n)|r(n)}=\rbraket{l(n-1)|\voperator{E}^{A(n)}_{A(n)}|r(n)}$, which we require to be one. Here, we represented the left and right density matrices as $D^2$-dimensional bra and ket vectors with round brackets via the Choi-Jamio{\l}kowski isomorphism \cite{Sudarshan:1961aa,Jamiokowski:1972aa,Choi:1975aa,Arrighi:2004aa}, and introduced the notation $\voperator{E}^A_B=\sum_{s=1}^{d} A^s \otimes \overline{B}^s$. For a left-canonical [right-canonical] MPS, where all the tensors are left orthonormalized [right orthonormalized], the left density matrices $l(n)$ [right density matrices $r(n)$] would be identity matrices. The matrices $C(n)$ that allow to go from one canonical form to the other according to $A^s_L(n) C(n) = C(n-1) A^s_R(n)$ can be used to compute the right density matrices $r(n)$ for the left-canonical form as $r(n) = C(n)C(n)^\dagger$, or the left density matrices for the right-canonical form as  $l(n)=C(n)^\dagger C(n)$. The eigenvalues of the density matrices are 
thus related to the square of the Schmidt coefficients. 

To introduce the MPS tangent space, it is required that the set of MPS constitutes a smooth manifold. It was indeed established that the set of full rank MPS ---meaning that all density matrices $l(n)$ and $r(n)$ as defined in the main text have full rank--- constitute a smooth manifold in Ref.~\onlinecite{Holtz:2012aa,Uschmajew2013133} (for real-valued MPS with open boundary conditions) and in Ref.~\onlinecite{Haegeman:2014aa} (for complex-valued finite MPS with open boundary conditions and uniform MPS with periodic boundary conditions or in the thermodynamic limit). Finite MPS for which some $l(n)$ and/or $r(n)$ don't have full rank, or uniform MPS which are not injective, correspond to singular points where the tangent space is not well defined. We briefly elaborate on this in the sections on the 2-site integrators. In the proofs of Refs.~\onlinecite{Uschmajew2013133,2012arXiv1210.7710H}, the mapping between the MPS parameters (the site dependent tensors $A(n)\in\mathbb{C}^{D_{n-1}\times d \times D_{n}}$) and the state $\ket{\Psi[A]}$ is identified as a principal fiber bundle, resulting from the gauge redundancy of MPS as discussed in the main text.

The MPS tangent $T_{\ket{\Psi[A]}} \mathcal{M}_{\text{MPS}}$ space is spanned by the partial derivatives of $\ket{\Psi[A]}$ with respect to all entries $A^s_{\alpha,\beta}(n)$ for every site $n=1,\ldots,N$. We now denote a general variation as $B^{i}$ with a collective index $i=(\alpha,s,\beta,n)$, such that the most general MPS tangen vector can be written as
\begin{equation}
B^{i}\ket{\partial_{i} \psi}=\sum_{n=1}^{N} \sum_{\{s_{n}\}=1}^{d} A^{s_{1}}(1) \cdots B^{s_{n}}(n) \cdots A^{s_{N}}(N)\ket{s_{1}\ldots s_{n}\ldots s_{N}}.\label{eq:mpstangent}
\end{equation}
The gauge redundancy of the MPS parameterization reflects itself in tangent space by the fact that not all linearly independent choices $B^i$ produce independent tangent vectors $B^i\ket{\partial_i\Psi[A]}$, i.e. the basis of partial derivatives is overcomplete. Put differently, the linear map from the parameters $B^i$ to the states $B^i\ket{\partial_i \Psi[A]}$ has a non-trivial kernel. In particular, the space of infinitesimal gauge transformations (also known as the vertical subspace) $B^{s}(n) = \mathcal{N}^s[X](n) = X(n-1) A^{s}(n) - A^{s}(n) X(n)$ with $X(n)\in\mathbb{C}^{D_n\times D_n}$ result in $B^{i} \ket{\partial_{i} \psi(A)}=0$, as can easily be checked by explicit insertion. Finding a unique parameterization for every tangent vector is equivalent to specifying a complement of the vertical subspace (the so-called horizontal subspace), for which there is no unique prescription. Within the framework of principal fibre bundles, a natural way to define the horizontal subspace is via a principal bundle connection. We simply refer to the result as a gauge fixing prescription for the tangent vectors and present two possible choices. One can describe any tangent vector by a parameterization $B$ that satisfies the `left gauge fixing condition'
\begin{align}
\rbra{l(n-1)}\voperator{E}^{B(n)}_{A(n)}=0,\qquad \forall n=1,\ldots,N-1\label{eq:tleft}
\end{align}
or alternatively the `right gauge fixing condition'
\begin{align}
\voperator{E}^{B(n)}_{A(n)}\rket{r(n)}=0,\qquad\forall n=2,\ldots,N.\label{eq:tright}
\end{align}
That these two gauge fixing conditions are related to a principal bundle connection implies that they transform covariantly under gauge transformations on the original MPS. If we transform $A^s(n)\to A^s_G(n)=G(n-1)^{-1} A^s(n) G(n)$ then $B^s(n)$ has to follow the identical transformation law to $B^s_G(n)=G(n-1)^{-1} B^s(n) G(n)$ in order to still satisfy the same gauge fixing condition. Extending the left gauge fixing condition to $n=N$ or the right gauge fixing condition to $n=1$ is equivalent to restricting to those variations of $\ket{\Psi[A]}$ that preserve norm and phase, \textit{i.e.}\ those tangent vectors which are orthogonal to $\ket{\Psi[A]}$. In combination with the left, respectively right orthonormal form for the MPS tensors, these gauge fixing conditions express that the orthonormality constraint is preserved to first order.

To derive the inverse-free action of the tangent space projector, we now switch to a different representation of tangent vectors, given by
\begin{equation}
\begin{split}
\ket{\Theta[B]}&=\sum_{n=1}^{N} \sum_{\{s_{n}\}=1}^{d} A_L^{s_{1}}(1) \cdots A_L^{s_{n-1}}(n-1) B^{s_{n}}(n) A_R^{s_{n+1}}(n+1)\cdots A_R^{s_{N}}(N)\ket{s_{1}\ldots s_{n}\ldots s_{N}}.\\
&= \sum_{n=1}^{N} \sum_{\{\alpha,\beta,s_n\}}  B^{s_n}_{\alpha,\beta}(n) \ket{\Phi^{[1:n-1]}_{L,\alpha}}\ket{s_n}\ket{\Phi^{[n+1:N]}_{R,\beta}}
\end{split}\label{eq:tangentnew}
\end{equation}
It is clear how this representation can be obtained from Eq.~\eqref{eq:mpstangent} by applying left orthonormalization and right orthonormalization in every term separately, and absorbing the resulting factors in the new definition of $B^{s_n}(n)$. In this new representation, however, there is no easy way to relate $B$ to the basis of partial derivatives. The vertical subspace is now given by choices $B^s(n) = \mathcal{N}^s[X](n) = X(n-1) A_R^{s}(n) - A_L^{s}(n) X(n)$ and allows to impose e.g.\ the `left gauge fixing condition' $\sum_{s} A_L^s(n) B^s(n) = 0$, $\forall n=1,\ldots,N-1$. This ensures that the overlap of two tangent vectors only contains diagonal contributions, as in
\begin{equation}
\braket{\Theta[B_1]|\Theta[B_2]}= \sum_{n=1}^{N} \sum_{s_n} \tr\left[ B_1^{s_n}(n)^\dagger B_2^{s_n}(n)\right].
\end{equation}
This corresponds to the standard Euclidean inner product and thus to the choice of an orthonormal basis. Note that imposing the `left gauge fixing condition` is essential, since the basis composed of the states
\begin{displaymath}
\left\{ \ket{\Phi^{[1:n-1]}_{L,\alpha}}\ket{s_n}\ket{\Phi^{[n+1:N]}_{R,\beta}}, \forall \alpha,\beta,s_n, \forall n=1,\ldots,N\right\}
\end{displaymath}
might seem orthonormal at first, but is not (consider the overlap between two states for different $n$) and is in fact overcomplete.

Let us now discuss action of the tangent space projector $P_{T_{\ket{\Psi[A]}\varM_{\text{MPS}}}}$ onto an arbitrary vector $\ket{\Xi}$ in the Hilbert space $\hilbert$. The resulting vector $P_{T_{\ket{\Psi[A]}\varM_{\text{MPS}}}}\ket{\Xi}$ should be an element of the tangent space $T_{\ket{\Psi[A]}\varM_{\text{MPS}}}$ of the manifold $\varM_{\text{MPS}}$ of MPS at the base point $\ket{\Psi[A]}$. It can thus be represented in the form $\ket{\Theta[B]}$. The standard Euclidean inner product of $\hilbert$ guarantees that the equation $\ket{\Theta[B]}= P_{T_{\ket{\Psi[A]}\varM_{\text{MPS}}}}\ket{\Xi}$ is also a solution of
\begin{equation}
\min_{B} \lVert \ket{\Theta[B]}-\ket{\Xi} \rVert^2\label{eq:projmin}.
\end{equation}
We can reformulate the minimization problem of Eq.~\eqref{eq:projmin} as
\begin{displaymath}
\min_{B} \sum_{n=1}^{N} \sum_{s} \big[B^s(n) B^s(n)^\dagger - B^s(n) F^s(n)^{\dagger}- F^s(n) B^s(n)^{\dagger}\big]
\end{displaymath}
together with the constraints
\begin{displaymath}
\sum_{s} A_L^s(n)^\dagger B^s(n)=0,\quad\forall n=1,\ldots,N-1.
\end{displaymath}
The tensor $F(n)$ can easily be read from omitting tensor $B(n)$ in the overlap between the $n$th term of Eq.~\eqref{eq:tangentnew} and the general state $\ket{\Xi}$, and corresponds to the overlap with the basis vectors 
\begin{equation}
F^{s_n}_{\alpha,\beta}(n)=\braket{\Phi^{[1:n-1]}_{L,\alpha},s_n,\Phi^{[n+1:N]}_{R,\beta}|\Xi}. 
\end{equation}
The solution to this constrained optimization problem is given by
\begin{displaymath}
B^s(n)=\sum_{t} (\delta_{s,t} - A_{L}^s(n) A_{L}^t(n)^{\dagger}) F^t(n)
\end{displaymath}
for all $n=1,\ldots,N-1$, where first factor is a projector that imposes the left-gauge fixing condition. Tensor $B(N)$ is not subject to any constraint and is given by $B^s(N)=F^s(N)$. For $n=1,\ldots,N-1$, we also introduce the notation $G(n)= \sum_{t} A_{L}^t(n)^{\dagger} F^t(n)$, or thus
\begin{equation}
G_{\alpha,\beta}(n)=\braket{\Phi^{[1:n]}_{L,\alpha},\Phi^{[n+1:N]}_{R,\beta}|\Xi},
\end{equation}
so that we can rewrite $B^s(n) = F^s(n) - A^s_L(n) G(n)$. Inserting this solution into $\ket{\Theta[B]}$ gives immediate rise to
\begin{equation}
\ket{\Theta[B]}=P_{T_{\ket{\Psi[A]}\varM_{\text{MPS}}}}\ket{\Xi}=\left(\sum_{n=1}^{N} \operator{P}_{L}^{[1:n-1]}\otimes \operator{1}_n\otimes \operator{P}_{R}^{[n+1:N]} -\sum_{n=1}^{N-1} \operator{P}_{L}^{[1:n]}\otimes \operator{P}_{R}^{[n+1:N]}\right)\ket{\Xi}
\end{equation}
where we can recognise the contents of the brackets as the decomposition of the tangent space projector into a sum of orthogonal projectors, where $\operator{P}_{L}^{[1:n]}$ and  $\operator{P}_{R}^{[n+1:N]}$ where defined in the main text.

This derivation differs from previous papers in using the new representation $\ket{\Theta[B]}$. Previous implementations of the TDVP required parameters that were directly defined with respect to the basis of partial derivatives $\ket{\partial_i\Psi[A]}$, which then specified how to update the parameters $A$ in the time-evolution. For the new representation $\ket{\Theta[B]}$, the parameters of $B$ cannot easily be related to basis of partial derivatives with respect to all parameters $A^s(n)$, $\forall n=1,\ldots,N$. The splitting scheme presented in the main text nevertheless allows us to update the MPS $\ket{\Psi}$ using this representation. 

\subsection{Full algorithm for the symmetric 1-site integrator}
\label{s:algo}
For completeness, and to indicate the strong similarity to a typical DMRG implementation, we here present the implementation of a full step of the symmetric 1-site integration scheme, which was presented in the main text and is based on a Lie-Trotter splitting scheme according to the decomposition of the tangent space projector as discussed in the previous section.

\begin{alg}[1-site symmetric integration step]
\label{alg:onesiteI}
Starting from a right-orthonormal MPS parameterization $\{A_C(1,t),A_R(2,t),\ldots,A_R(N,t)\}$ for the state at time $t$, compute a right-orthonormal MPS parameterization $\{A_C(1,t+\Delta t),A_R(2,t+\Delta t),\ldots, A_R(N,t+\Delta t)\}$ for the state at time $t+\Delta t$ by applying the following steps:
\begin{enumerate}
\item Repeat for n=1:N-1 \# left-to-right sweep
\begin{enumerate}
\item Evolve $A_C(n,t)$ forward in time according to $$\bm{A}_C(n,t+\Delta t/2)=\exp[-\ic H(n) \Delta t/2] \bm{A}_C(n,t)$$
where $H(n)$ is computed using $A_L(1,t+\Delta t/2)$, \ldots, $A_L(n-1,t+\Delta t/2)$, $A_R(n+1,t)$, \ldots, $A_R(N,t)$.
\item Perform an orthogonal decomposition of $A_C(n,t+\Delta t/2)$ into $A_L(n,t+\Delta t/2)$ and $\tilde{C}(n,t+\Delta t/2)$.
\item Evolve $\tilde{C}(n,t+\Delta t/2)$ backwards in time according to $$\bm{\tilde{C}}(n,t)=\exp[+\ic K(n) \Delta t/2] \bm{\tilde{C}}(n,t+\Delta t/2)$$ where $K(n)$ is computed using $A_L(1,t+\Delta t/2)$,\ldots,$A_L(n,t+\Delta t/2)$, $A_R(n+1,t)$, \ldots, $A_R(N,t)$.
\item Absorb $\tilde{C}(n,t)$ into $A_R(n+1,t)$ to obtain $A_C(n+1,t)$.
\end{enumerate}
\item Evolve $A_C(N,t)$ forward in time according to $$\bm{A}_C(N,t+\Delta t)=\exp[-\ic H(N) \Delta t] \bm{A}_C(N,t)$$
where $H(N)$ is computed using $A_L(1,t+\Delta t/2)$, \ldots, $A_L(N-1,t+\Delta t/2)$.
\item Repeat for n=N-1:-1:1 \# right-to-left sweep
\begin{enumerate}
\item Perform an orthogonal decomposition of $A_C(n+1,t+\Delta t)$ into $\tilde{C}(n,t+\Delta t)$ and $A_R(n,t+\Delta t)$.
\item Evolve $\tilde{C}(n,t+\Delta t)$ backwards in time according to $$\bm{\tilde{C}}(n,t+\Delta t/2)=\exp[+\ic K(n) \Delta t/2] \bm{\tilde{C}}(n,t+\Delta t)$$ where $K(n)$ is computed using $A_L(1,t+\Delta t/2)$,\ldots,$A_L(n,t+\Delta t/2)$, $A_R(n+1,t+\Delta t)$, \ldots, $A_R(N,t+\Delta t)$.
\item Absorb $\tilde{C}(n,t+\Delta t/2)$ into $A_L(n,t+\Delta t/2)$ to obtain $A_C(n+1,t+\Delta t/2)$.
\item Evolve $A_C(n,t+\Delta t/2)$ forward in time according to $$\bm{A}_C(n,t+\Delta t)=\exp[-\ic H(n) \Delta t/2] \bm{A}_C(n,t+\Delta t/2)$$
where $H(n)$ is computed using $A_L(1,t+\Delta t/2)$, \ldots, $A_L(n-1,t+\Delta t/2)$, $A_R(n+1,t+\Delta t)$, \ldots, $A_R(N,t+\Delta t)$.
\end{enumerate}
\end{enumerate}
\end{alg}

Let us briefly elaborate on the philosophy behind this integration scheme. If we separate the evolution in step 2 into two consecutive evolutions with time step $\Delta t/2$, referred to as step 2(a) and 2(b) respectively, then one can recognize the right-to-left sweep [step 2(b) and everything in step 3] as the adjoint method to the left-to-right sweep [everything in step 1 combined with step 2(a)]. For a general finite-step integrator $\mathcal{I}_{\Delta t}$ with time step $\Delta t$, the adjoint method $\mathcal{I}^{\ast}_{\Delta t}$ is defined as the inverse map of $\mathcal{I}_{-\Delta t}$. The combined scheme $\mathcal{I}_{\Delta t/2}^{\ast} \circ \mathcal{I}_{\Delta t/2}$ is symmetric by construction. If $\mathcal{I}_{\Delta t}$ is a first order integrator (meaning that the local error scales as $\mathcal{O}(\Delta t^2)$, the combined scheme becomes second order. However, it is not often the case that the adjoint method $\mathcal{I}^{\ast}$ can exactly be evaluated. For example, the adjoint method of the explicit Euler method is the implicit Euler method. That the adjoint method for the finite MPS integrator can be evaluated explicitly can be traced back to the fact that the individual terms obtained from splitting the tangent space projector are exactly integrable. The underlying reason for this is that the effective Hamiltonian $H(n)$ [or $C(n)$] for the 1-site [or 0-site] center degrees of freedom $A_C(n)$ [or $C(n)$] does not depend on these degrees of freedom, but only on the tensors on the neighbouring sites, so that the differential equations corresponding to the individual terms are first-order linear differential equations.

\subsection{Two-site integration scheme}
For general variational manifold, the TDVP principle of projecting the evolution vector onto the tangent space is the only strategy that ensures that the evolution can be described within the set of states that are assumed tractable, i.e.\ the manifold itself. Matrix product states as variational ansatz are special in the sense that they have (a set of) refinement parameters corresponding to the virtual dimensions. There is thus not a single variational manifold, but rather a hierarchy of manifolds, where the ones with lower bond dimension actually correspond to the singular regions of the ones with higher bond dimension. This raises the question whether we can generalize the `TDVP philosophy' so as to obtain an evolution within this hierarchy of states. From DMRG and TEBD, we know that such an evolution typically involves acting on a center block of two sites. The two-site evolution will take the state out of the manifold of MPS with fixed bond dimensions. While one typically truncates back to the original manifold, the advantage is that such an approach allows to truncate to a different manifold with higher (or lower) bond dimension when this is still computationally feasible. For the remainder, we assume that the truncation scheme is based on discarding singular values smaller than some tolerance level $\varepsilon$. Unfortunately, such a strategy cannot easily be formulated as a differential equation, as it necessarily requires a finite time step $\Delta t$. We can nevertheless get inspired by the normal TDVP derivation.

We start by defining the linear space of two-site variations to the current MPS $\ket{\Psi[A]}$. This is the linear space contains states of the form
\begin{equation}
\sum_{n=1}^{N-1} \sum_{\{s_{n}\}=1}^{d} A_L^{s_{1}}(1)  \cdots A_L^{s_{n-1}}(n-1)B^{s_{n}s_{n+1}}(n:n+1) A_R^{s_{n+2}}(n+2) \cdots A_R^{s_{N}}(N)\ket{s_{1}\ldots s_{n}\ldots s_{N}}.
\label{eq:states2}
\end{equation}
Let us call this space $T^{[2]}_{\ket{\Psi[A]}} \varM_{\mathrm{MPS}}$. Note that this notation should not be confused with the double tangent space, which contains the span of all single-site variations on two arbitrary sites. As for the tangent space, the linear map from the representation $\{B^{s,t}(n:n+1);n=1,\ldots,N-1\}$ to the physical states in $T^{[2]}_{\ket{\Psi[A]}} \varM_{\mathrm{MPS}}$ has a non-trivial kernel corresponding to choices
\begin{displaymath}
B^{s,t}(n:n+1)=A_L^s(n) X^t(n+1)-X^s(n) A_R^t(n+1)
\end{displaymath}
for any $\{X(n)\in\mathbb{C}^{D_{n}\times d\times D_{n+1}},n=2,\ldots,N-1\}$. This freedom allows to impose a gauge-fixing condition of the form
\begin{equation}
\sum_{s} A_L^s(n)^\dagger B^{s,t}(n:n+1)=0,\quad \forall n=1,\ldots,N-2.
\end{equation}
The orthogonal projector $\operator{P}_{T^{[2]}_{\ket{\Psi[A]}} \varM_{\mathrm{MPS}}}$ that maps any state $\ket{\Phi}\in\hilbert$ onto the subspace $T^{[2]}_{\ket{\Psi[A]}} \varM_{\mathrm{MPS}}$ can then be constructed and decomposed ---using manipulations similar to those of the previous subsection--- as a sum of orthogonal projections
\begin{equation}
\operator{P}_{T^{[2]}_{\ket{\Psi[A_L]}}\varM_{\text{MPS}}} = \sum_{n=1}^{N-1} \operator{P}_{L}^{[1:n-1]}\otimes \operator{1}_n\otimes \operator{1}_{n+1}\otimes \operator{P}_{R}^{[n+2:N]}-\sum_{n=2}^{N-1} \operator{P}_{L}^{[1:n-1]}\otimes\operator{1}_{n} \otimes \operator{P}_{R}^{[n+1:N]}
\label{eq:proj2}
\end{equation}
In particular, when applying $\operator{P}_{T^{[2]}_{\ket{\Psi[A_L]}}\varM_{\text{MPS}}}$ onto $\operator{H}\ket{\Psi[A]}$, one obtains
\begin{equation}
\begin{split}
\operator{P}_{T^{[2]}_{\ket{\Psi[A_L]}}\varM_{\text{MPS}}}\operator{H}\ket{\Psi[A]}= &\sum_{n=1}^{N-1}\sum_{\alpha,\beta,s_n,s_{n+1}} B^{s_n s_{n+1}}_{\alpha,\beta}(n:n+1) \ket{\Phi^{[1:n-1]}_{L,\alpha},s_n,s_{n+1},\Phi^{[n+2:N]}_{R,\beta}} \\
&-\sum_{n=2}^{N-1}\sum_{\alpha,\beta,s_n,s_{n+1}} B^{s_n}_{\alpha,\beta}(n) \ket{\Phi^{[1:n-1]}_{L,\alpha},s_n,\Phi^{[n+2:N]}_{R,\beta}}
\end{split}\label{eq:proj2ham}
\end{equation}
where $\bm{B}(n)=H(n) \bm{A}_{C}(n)$ and now also $\bm{B}(n:n+1)=H(n:n+1) \bm{A}_{C}(n:n+1)$, with $H(n)$ and $H(n:n+1)$ the effective one-site and two-site Hamiltonians defined in the context of DMRG in the main text. This result justifies the 2-site integration scheme that was proposed in the main text.

The error analysis of this 2-site integrator is more complicated, as there does not exist an underlying differential equation that describes the evolution projected down to $T^{[2]}_{\ket{\Psi[A_L]}}\varM_{\text{MPS}}$. With respect to the full Schr\"{o}dinger equation, we except a competition between the finite time step error, a truncation error $\varepsilon$, and a projection error that can be estimated in terms of $\lVert (\operator{1}-\operator{P}_{T^{[2]}_{\ket{\Psi[A_L]}}\varM_{\text{MPS}}})\operator{H}\ket{\Psi[A]}\rVert$. Without truncation ($\varepsilon=0$), one can easily verify that, after applying the complete sweep (left-to-right and back) with a time step $\Delta t$ to an MPS $\ket{\Psi[A]}$, the original physical state $\ket{\Psi[A]}$ is obtained by reapplying the scheme with a time step $-\Delta t$, although $\ket{\Psi[A]}$ will now be encoded as an MPS with a larger bond dimension (and matrices $l(n)$ or $r(n)$ which don't have full rank). Hence, without the truncation, the integrator is still symmetric and one expects a finite time step error that scales as $\order(\Delta t^4)$. This should still hold for $\epsilon>0$, since any error of lower order in $\Delta t$ should scale at least proportional to $\epsilon$ and is thus negligible in comparison to the contribution of $\order(\epsilon)$ that is introduced by the truncation anyway.  Instead of truncating after every step in the algorithm, we could also find the best MPS approximation with lower bond dimension after a complete sweep without truncation \cite{2004PhRvL..93t7204V}.

In the case of a nearest neighbor Hamiltonian, the projector error is absent as the action of $\operator{H}$ on $\ket{\Psi[A]}$ is exactly contained in $T^{[2]}_{\ket{\Psi[A_L]}}\varM_{\text{MPS}}$. It is then also possible to apply a Lie-Trotter splitting scheme to the individual terms in the Hamiltonian. This is exactly the scheme that gives rise to the TEBD algorithm. In contrast, our scheme is based on the splitting of the projector in Eq.~\eqref{eq:proj2} and can be applied to arbitrary Hamiltonians (for which $H(n)$ and $H(n:n+1)$ can be computed efficiently) with the same computational cost. The difference between both schemes can be related to redundancy in the representation of states in $T^{[2]}_{\ket{\Psi[A_L]}}\varM_{\text{MPS}}$. In particular, by directly applying a nearest neighbor Hamiltonian $\operator{H}=\sum_{n=1}^{N-1} \operator{h}(n:n+1)$ to $\ket{\Psi[A]}$, one obtains a state of the form of Eq.~\eqref{eq:states2} which is represented by $B^{s_1,s_2}(n:n+1)=\sum_{t_1,t_2} \braket{s_1,s_2|\operator{h}(n:n+1)|t_1,t_2} A^{s_1}(n) A^{s_2}(n)$. This is different from the representation in Eq.~\eqref{eq:proj2ham}, even though both describe the same physical state $\operator{H}\ket{\Psi[A]}$. 

\end{widetext}

\end{document}